# Anomalous Nernst effect on the nanometer scale: Exploring three-dimensional temperature gradients in magnetic tunnel junctions

Ulrike Martens[1], Torsten Huebner[2], Henning Ulrichs[3], Oliver Reimer[2], Timo Kuschel[2], Ronnie R Tamming[4], Chia-Lin Chang[4], Raanan I Tobey[4], Andy Thomas[5], Markus Münzenberg[1], Jakob Walowski[1]

1 Institut für Physik, Ernst-Moritz-Arndt Universität Greifswald, Felix-Hausdorff-Straße 6, 17489 Greifswald, Germany

2 Center for Spinelectronic Materials and Devices, Physics Department, Bielefeld University, Universitätsstraße 25, 33615 Bielefeld, Germany

3 I. Physikalisches Institut, Georg-August-Universität Göttingen, Friedrich-Hund-Platz 1, 37077 Göttingen, Germany

4 Optical Condensed Matter Physics, Zernike Institute for Advanced Materials, University of Groningen, Nijenborgh 4, 9747 AG Groningen, The Netherlands

5 Leibniz Institute for Solid State and Materials Research Dresden (IFW Dresden), Institute for Metallic Materials, Helmholtzstrasse 20, 01069 Dresden, Germany

E-Mail: jakob.walowski@uni-greifswald.de, ulrike.martens@uni-greifswald.de

**Abstract.** Localized laser heating creates temperature gradients in all directions and thus leads to three-dimensional electron flux in metallic materials. Temperature gradients in combination with material magnetization generate thermomagnetic voltages. The interplay between these direction-dependent temperature gradients and the magnetization along with their control enable to manipulate the generated voltages, e.g. in magnetic nanodevices. We identify the anomalous Nernst effect (ANE) generated on a nanometer length scale by micrometer sized temperature gradients in magnetic tunnel junctions (MTJs). In a systematic study, we extract the ANE by analyzing the influence of in-plane temperature gradients on the tunnel magneto-Seebeck effect (TMS) in three dimensional devices. To investigate these effects, we utilize in-plane magnetized MTJs based on CoFeB electrodes with an MgO tunnel barrier. Due to our measurement configuration, there is no necessity to disentangle the ANE from the spin Seebeck effect in inverse spin-Hall measurements. The temperature gradients are created by a tightly focused laser spot. The spatial extent of the measured effects is defined by the MTJ size, while the spatial resolution is given by the laser spot size and the step size of its lateral translation. This method is highly sensitive to low voltages and yields an ANE coefficient of $K_\mathrm{N} \approx 1.6 \cdot 10^{-8} \frac{V}{TK}$ for CoFeB. In general, TMS investigations in MTJs are motivated by the usage of otherwise wasted heat in magnetic memory devices for read/write operations. Here, the additionally generated ANE effect allows to expand the MTJs' functionality from simple memory storage to nonvolatile logic devices and opens new application fields such as direction dependent temperature sensing with the potential for further downscaling.

# 1. Introduction

Spin-dependent thermally driven transport phenomena have the potential to expand the functionality of today's conventional electronics. A dream of spintronic researchers has been to improve not solely the devices speed, but also enhance power management. This can be accomplished by employing additional energy conversion mechanisms usually available in semiconductor based integrated circuitry in the form of waste heat. The emerging field of spin caloritronics takes advantage of spin electronic devices in combination with thermal effects. This research field stands at the frontier between thermal transport and spin physics [1–3]. Magnetic tunnel junctions (MTJs) are one great testbed for spin caloritronic application devices. Originally, they were developed for storage capacity enhancement by the use of the tunnel magnetoresistance effect (TMR) [4]. However, their properties can be directly translated to spin caloritronics, when an electric potential as a driving force is replaced by temperature gradients. The thermal method generate voltage and read out information from MTJs employing temperature gradients utilizes the tunnel magneto-Seebeck effect (TMS). When a temperature gradient is applied across a layer stack of two magnetic electrodes separated by an insulating barrier, the generated voltage $V$ differs, depending on whether the electrodes' magnetizations are aligned parallel (p) or antiparallel (ap). The microscopic origin together with theoretical predictions of the TMS for multiple CoFe compositions with MgO barriers is given in reference [5, 6] and is calculated by:

$$\text{TMS} = \frac{V_{\text{ap}} - V_{\text{p}}}{\min(|V_{\text{ap}}|, |V_{\text{p}}|)}.$$

*(1)*

The TMS effect has been observed and analyzed for various combinations of barrier and electrode materials, showing thermovoltages in the μV range for MgO [7–13] and $MgAl_2O_4$ [14, 15], and reaching the mV range for Heusler based MTJs [16]. All examined material configurations result in specific TMS ratios. While $MgAl_2O_4$ exhibits ratios below 10%, MgO reaches values up to 60% [15] and for electrode combinations CoFeB/MgO/Heusler even ratios of approximately 100% are reported. Meanwhile, the thermal voltage amplitudes approach the order of magnitude that could be used in commercial electronics. Besides this, other effects e.g. the Onsager reciprocal effect, the tunnel Peltier effect has been realized experimentally [17].

Two preconditions are required to unambiguously achieve enhanced Seebeck voltages in the MTJ's parallel and antiparallel state, $V_{\text{p}}$ and $V_{\text{ap}}$, [9, 11]. One must apply a large temperature gradient that across the junction and the whole junction area needs to be heated homogeneously. As a consequence, using all-optical laser heating, the spot size needs to be adapted to the junction size and positioned centrally in order to create a well-defined temperature gradient across both electrodes and generate reliable voltages [11]. Temperature gradients deviating from the out-of-plane direction, for example temperature inhomogeneities in the sample plane, lead to additional thermoelectric effects that influence the total Seebeck voltages. In this study we focus on effects generated by these in-plane temperature gradients.

There are three thermomagnetic effects that come into question when considering ferromagnetic metal materials whose temperature gradient $\nabla T$ and the magnetization $M$ are aligned in the same plane. The first two are the anisotropic magneto thermopower (AMTP) $E_{\text{AMTP}} \propto \nabla T \cdot \cos(\phi_{\nabla T}) \cdot M^2 \cdot \cos(2\phi_M)$ and the planar Nernst effect (PNE) $E_{\text{PNE}} \propto \nabla T \cdot \sin(\phi_{\nabla T}) \cdot M^2 \cdot \sin(2\phi_M)$. The angles $\phi_{\nabla T}$ and $\phi_M$ express the direction of $\nabla T$ and $M$ with respect to the direction of voltage measurement. The third is the anomalous Nernst effect (ANE), $E_{\text{ANE}} \propto \nabla T \times M \propto \nabla T \cdot$ $M \cdot \sin(\phi)$. The angle $\phi$ denotes the angle between the magnetization $M$ and the temperature gradient $\nabla T$. In the first two configurations, the generated electric fields $E_{\text{AMTP}}$ and $E_{\text{PNE}}$ are both coplanar with $\nabla T$ and $M$ [18], while in the last case, the voltage is orthogonal to both, $\nabla T$ and $M$. The former two effects are quadratic in $M$, which means, that magnetization reversal (180° rotation) would not lead to any change in the voltage direction. For the ANE, the resulting electric field is perpendicular to the plane defined by $\nabla T$ and the $M$ vector, and in contrast to the former two it exhibits a sign change upon magnetization reversal. In general, ANE experiments are performed with an out-of-plane temperature gradient and the magnetization in the film plane (IM configuration), as published in references [19–28], as well as with in-plane temperature gradients and perpendicular magnetization (PM configuration), see references [24, 29, 30]. In those experiments, usually macroscopic millimeter sized structures and micrometer wide wires are investigated. The voltage is generated on macroscopic length scales ranging from $> 10$ µm to several millimeters probing predominantly bulk-like properties. In this scope, the ANE measurements in perpendicular magnetized CoFeB nanowires with thicknesses below 1 nm play a special role, because those are the smallest dimensions, in which the ANE has been reported so far. There, the temperature gradients are created on length scales up to 500 nm and the generated voltages are detected on length scales in the micrometer range [31, 32].

In the present study, we utilize an extended TMS measurement configuration to deliberately create in-plane temperature gradients in MTJ electrodes with in-plane magnetization easy axis and detect the underlying thermomagnetic processes on mesoscopic length scales. This is done by the application of complex three-dimensional temperature gradients to drive spin caloritronic effects in the layered device. We exploit a high flexibility to control both, the magnetization and the temperature gradient direction, and measure the voltages in the lithographically structured MTJ.

In our experimental configuration the voltage is measured in the out-of-plane direction (defined as z-axis), perpendicular to the applied magnetic field $\mu_0 H$ (defined as y-axis) while the in-plane temperature gradient is rotated in the x-y plane (see figure 1).

We use pseudo-spin valves, because of their most simplistic layer structure and possibility to control the magnetization in both layers of the MTJs. In contrast to exchange biased spin valves, where one magnetic layer is pinned, these devices allow multiple magnetic configurations in the parallel magnetization alignment: With respect to the temperature gradient the magnetization of both ferromagnetic electrodes can be rotated together. We define the thermovoltage measured in the parallel state for direction 1 and 2 as $V_{\text{p1}}$ and $V_{\text{p2}}$, as depicted in figure 2. The difference $\Delta V_{\text{ANE}} = V_{\text{p1}} - V_{\text{p2}}$ is employed in the following to disentangle and characterize thermomagnetic effects that arise from in-plane temperature gradients created in the plane of the electrodes.

The structure we use is briefly outlined as follows. First, the sample geometry and layer sequence are sketched to explain temperature distribution within the corresponding geometries and the applied external magnetic fields. The access to temperature gradients and the relevant temperature differences is discussed by finite element simulations using COMSOL. Second, the data extraction procedure is given together with the analysis, from which we conclude the presence of the ANE in our experiments. After that, additional experimental data is presented that identifies the uniaxial magnetic anisotropy present in the investigated samples. In the end, we compare the ANE constant to findings from other experimental techniques and discuss new possible applications for this effect based on micro- and nanoscale MTJ devices.

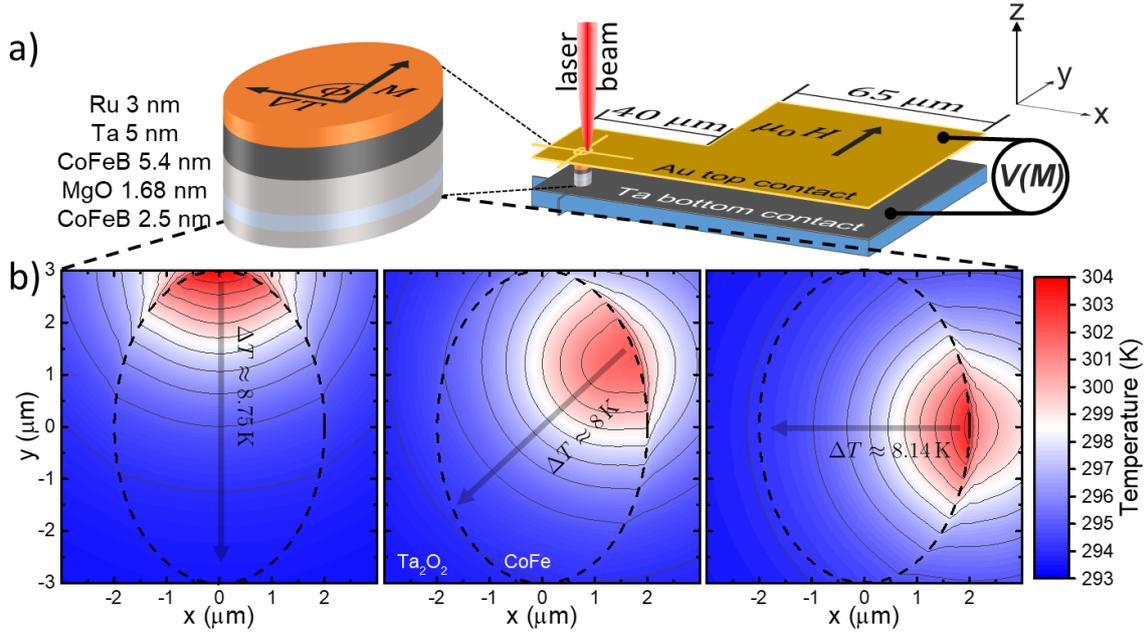

*Figure 1* **a)** Left: The MTJ layer stack with the corresponding thicknesses. The magnetization $M$, with respect to the in-plane temperature gradient $\nabla T$ (black arrows) their relation is given by the in-pane varied angle $\phi$. Right: MTJ between the Au top and the Ta bottom contacts. The direction of the incoming laser beam, the external magnetic field $\mu_0 H$ and the thermovoltage measurement configuration are indicated within the coordinate axes. **b)** False color plots showing the in-plane temperature gradients in the top CoFeB electrode for three heating laser spot positions at the junction edges obtained from COMSOL simulations. The MTJ areas are indicated by the dashed lines. Heating at the end of the long (short) axis results in the main temperature gradient in y-direction (x-direction) as shown in the left (right) graph and is indicated by the black arrows. The middle graph shows the scenario, when the heating laser spot is placed on the edge between both major ellipses axes, resulting in the main temperature gradient at an angle between the two.

## 2. Temperature distribution

The key feature to analyze spin caloritronic effects in MTJs is the access to temperature distributions on micrometer to nanometer length-scales. TMR junctions provide a rich variety of possibilities to create anisotropic temperature profiles on nm to µm length-scales using position-dependent laser heating. Extending the scanning technique originally developed for the extraction of the preferably pure TMS signal, as introduced in reference [11], allows a systematic temperature gradient variation. The schematic in Figure 1a) depicts this experimental procedure. In general, a centrally positioned laser spot creates a temperature gradient through the layer stack in z-direction that generates a magnetization dependent voltage $V(M)$ which can be varied by sweeping an external magnetic field $\mu_0 H$ applied in y-direction. The MTJ layer stack itself is embedded into Au and Ta contact pads and surrounded by insulating $Ta_2O_5$ in the x-y-plane. The Au pads thickness is around three times larger than the optical penetration depth $\lambda_{opt} \approx 15 - 20 nm$, leaving purely thermal excitation in the CoFeB layers. This sample design allows to create and steadily vary temperature gradients in the x-y-plane by moving the laser spot along the surface. The voltage generated at the CoFeB electrodes is measured in z-direction. Due to this configuration the main voltage contribution generated by in-plane temperature gradients stems from the ANE. Both the AMTP and the PNE can be disregarded, because the voltage is generated in the x-y-plane, and only second order processes with amplitudes that are orders of magnitude smaller can contribute to the out-of-plane signal. For the investigation of inhomogeneous laser heating, the setup parameters need additional adjustment. The modulated continuous wave laser spot is focused down to 2 µm in diameter and systematically scanned across the sample within an area of $30 \times 30$ µm² in which the elliptically shaped MTJ itself has a dimension of 6 µm by 4 µm. Performing such a 2-dimensional scan, a local heating point is moved over the entire MTJ area, and enables the creation of specifically directed and consistently varied temperature gradients. This allows us to apply complex three-dimensional temperature profiles at will. The situation is discussed in the following example when we place the laser spot at the MTJ's edge.

Figure 1a) (left) shows an enlargement of the elliptically shaped MTJ layer stack. The tunnel junction consists of the CoFeB/MgO/CoFeB stack, the Ta layer is necessary to remove boron during crystallization from the CoFeB/MgO interface and lastly the Ru capping is deposited to prevent oxidation during the ex-situ annealing process and patterning. The in-plane $\nabla T$ together with the in-plane $M$ and the angle $\phi$ are sketched on top of the stack. Note, that during the measurement, the direction of $M$ remains constant, while $\nabla T$ is rotated by $\phi = 0° - 360°$. The access to temperature in such small devices is not available experimentally, therefore, three-dimensional finite element simulations using the COMSOL package with the heat transfer module are performed to gain insight into the temperature distribution within the MTJ. The simulations are performed for continuous wave laser heating in equilibrium.

Figure 1b) displays the temperature distribution for three different laser spot positions located at the MTJ edges. The false color plots show the equilibrium temperature distribution inside the top CoFeB electrode which is indicated by the dashed lines. The temperature distribution in the bottom CoFeB shows the same characteristics and is not shown here. However, due to $\nabla T$ created in the out-of-plane direction, the overall $T$ is slightly lower. The difference in temperature between top and bottom electrodes ranges from $\Delta T_{\text{top-bottom}} \approx 50$ mK in the vicinity of the laser spot to less than 1 mK at the opposite edge and decreases exponentially. Since the dimensions in the x-y-plane are three orders of magnitude larger, the in-plane temperature differences are larger than those across the layer stack. The temperature gradient directions for each heating scenario are indicated by the gray arrows accompanied by the temperature drop $\Delta T$ between both junction edges.

The left plot describes the first scenario, when the laser spot is located at the vertex, then a temperature gradient along the major-axis with a temperature difference $\Delta T \approx 9$ K is created. The right plot illustrates the second scenario, when the laser spot is located at the co-vertex, resulting in a temperature gradient along the minor-axis with $\Delta T \approx 8$ K. Finally, the middle plot shows the third heating scenario when the laser spot is located at the edge of the ellipse at a 45° angle between both principal axes. Consequently, this results in a temperature gradient along the MTJ diagonal with $\Delta T \approx 8$ K. The slight temperature differences for these extremal cases results from the asymmetry in the MTJ's geometry. A thorough temperature profile analysis reveals that independent of the $\nabla T$ angle the in-plane temperature gradient covers equally sized areas of the MTJ. Therefore, we expect the number of electrons involved in the process triggered by the in-plane temperature gradient to remain angle independent.

The largest, most homogeneous area with a high temperature gradient across the layer stack is created when the laser spot is located with its center at least 1.7 µm away from the MTJ's edge. When heating within this area, effects from in-plane temperature gradients can be excluded. We conclude that by application of the laser spot at the edge of the tunnel junction, large in-plane gradients can be created that we can rotate by an arbitrary angle in the x-y-plane that dominate the three-dimensional temperature distribution. For a laser spot at the center, the overall x-y-gradient is found to vanish, and we have predominantly a temperature gradient in z-direction.

## 3. Anomalous Nernst effect

The pseudo spin valves selected in this study allow for the full directional manipulation of the magnetization in both electrodes, because in contrast to conventional MTJ design, none of the magnetic layers is antiferromagnetically pinned. The condition for their antiparallel magnetization alignment is realized by choosing electrodes with different anisotropy strength and thus different coercive fields. In the presented investigation both CoFeB layer differ in thicknesses by around 2 nm to fulfil this criterion. This allows two parallel magnetization alignment configurations of opposite direction.

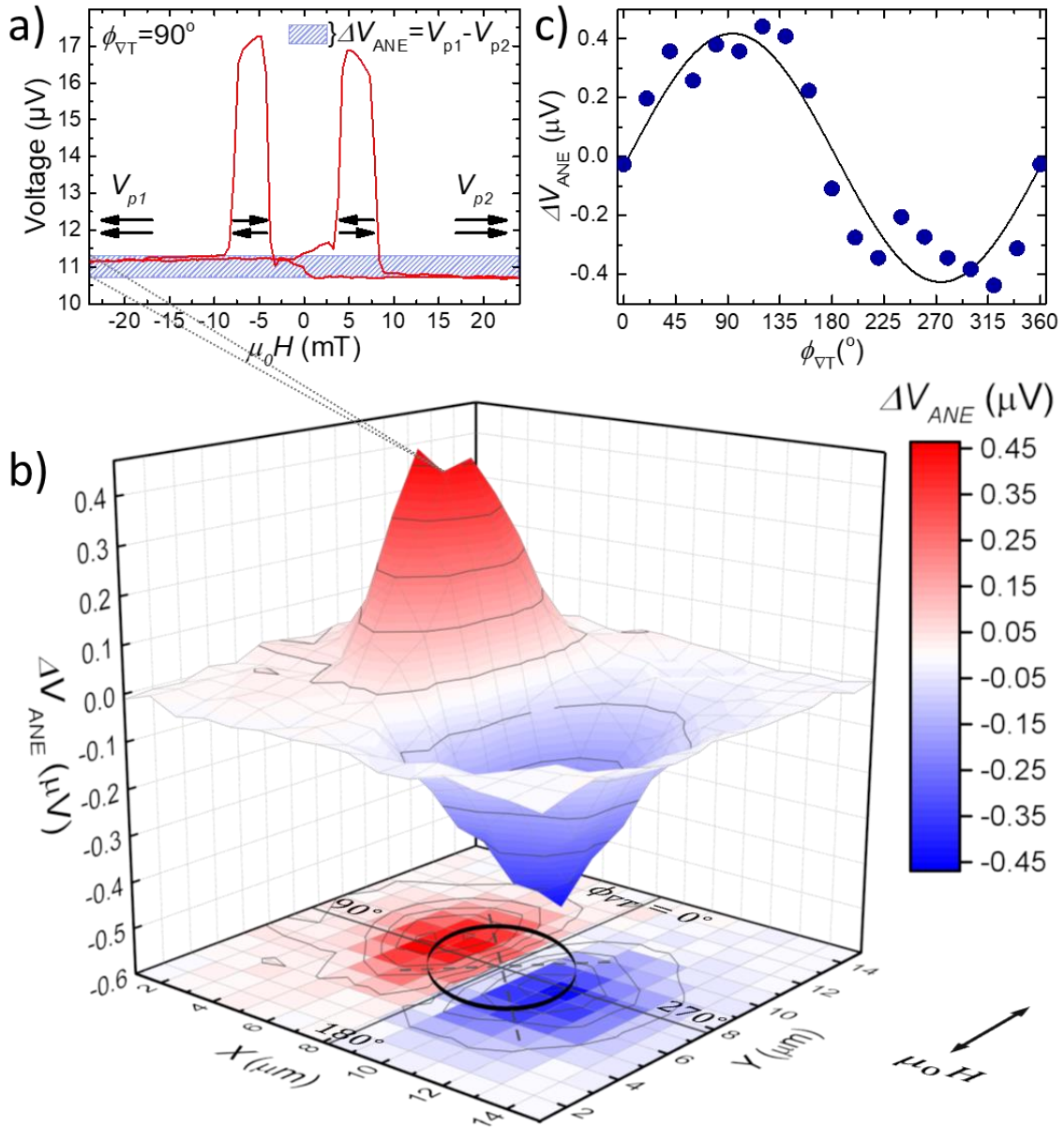

*Figure 2* **a)** Exemplary TMS measurement curve showing the Seebeck voltage vs. the external field (red line). The ranges with parallel and antiparallel magnetization alignment of both electrodes are indicated by the black arrows. The blue area marks the difference $\Delta V_{\mathrm{ANE}}$ between the voltage measured in parallel magnetization configuration for both directions $V_{p1}$ and $V_{p2}$. **b)** The extracted $\Delta V_{\mathrm{ANE}}$ values are plotted vs. the laser position in a three-dimensional surface plot with a false color projection at the bottom. The black ellipse outlines the MTJ area, with the grey cross located along the principal axes. **c)** $\Delta V_{\mathrm{ANE}}$ values extracted along the ellipses outline plotted against the temperature gradient angle $\phi_{\nabla T}$ together with the fitted sine curve.

Figure 2 a) shows an example of a Seebeck voltage vs. external field $\mu_0 H$ sweep, recorded while the laser spot is close to the MTJ edge and a pronounced in-plane temperature gradient is generated, $\phi_{\nabla T} = 90°$. The ranges with parallel and antiparallel magnetization alignment are indicated by the black arrows. For large field amplitudes, both electrodes magnetizations align parallel and a different Seebeck voltage is generated than in the antiparallel alignment. From this measurement curve, the TMS ratio is calculated, which results in a ratio of approximately 50 %. This is consistent with the findings reported in reference [11].

The measurement confirms the two possibilities for parallel alignment configuration of opposite direction, $V_{p1}$ for negative $\mu_0 H$ and $V_{p2}$ for positive $\mu_0 H$. Furthermore, the data exhibits a clear shift of $V_{p1}$ with respect to $V_{p2}$. This voltage shift $\Delta V_{ANE} = V_{p1} - V_{p2}$ is marked by the shaded blue area. We argue that $\Delta V_{ANE}$ originates from the in-plane temperature gradient, which affects the voltage in the perpendicular direction for parallel magnetization states of opposite sign. At this point we rule out the PNE and the AMTP. Their quadratic dependence on the magnetization $\sim M^2$ would not result in a difference between $V_{p1}$ and $V_{p2}$ upon magnetization reversal. Conclusively we state that $\Delta V_{ANE}$ originates from the ANE.

The laser spot is moved over the sample surface and the in-plane temperature gradient is varied, as analyzed in the previous section from the finite element temperature simulations. From each curve, one $\Delta V_{ANE}$ value is extracted and plotted in figure 2 b). In each measurement, the magnetization is reversed together with $\mu_0 H$ along the y-axis, as depicted by the black double arrow next to the graph. Figure 2 b) is divided into two parts.

In the first part, the extracted $\Delta V_{ANE}$ values for each heating scenario are illustrated in a three-dimensional surface plot. The spatial position for the $\Delta V_{ANE}$ value extracted from figure 2 a) is indicated by the gray dotted lines pointing to figure 2 b). The voltage difference $\Delta V_{ANE}$ shows an increase and a decrease with absolute value maxima of around 0.4 µV showing an inversion symmetry regarding the origin of the coordinate system.

In the second part, the same data is projected at the bottom in a false color plot for a better overview. This depiction includes an outline of the MTJ's elliptical area with both principal axes (dashed dark gray crossed lines). Without loss of generality, the angle $\phi_{\nabla T} = 0°$ is defined along the positive y-axis and parallel to the positive $\mu_0 H$ and the $\phi_{\nabla T}$ rotation is marked in counter clock-wise direction. Both extreme values of $\Delta V_{ANE}$ are generated when the laser heating spot is located near the MTJ's edge, where the largest in-plane temperature differences are created (compare COMSOL simulations in figure 1) and at $\phi_{\nabla T} = 90°$ and $\phi_{\nabla T} = 270°$. The borderline between the elevation and decrease where $\Delta V_{ANE} \approx 0$ proceeds parallel to $\mu_0 H$ and is perpendicular to the line connecting the extreme $\Delta V_{ANE}$ absolute value locations.

As a main result, the $\Delta V_{ANE}$ values extracted from the positions marked by the black ellipse outline are plotted versus the temperature gradient angle $\phi_{\nabla T}$ with respect to the $\mu_0 H$ direction is shown in figure 2 c). This two-dimesional plot highlights the $\Delta V_{ANE}$ sign change upon in-plane $\nabla T$ reversal with respect to the magnetization. This behavior confirms the thermomagnetic origin of the extracted effect. Further analysis of the $\Delta V_{ANE}$ signal in figure 2 c) validates the ANE effect generated by in-plane temperature gradients. The extracted data (blue dots) are fitted to the formula given by the ANE cross product definition, when the temperature gradient is rotated by $\phi_{\nabla T}$:

$$\Delta V_{ANE} = A \cdot \sin(\phi_{\nabla T} - \phi_0) + V_0.$$

*(2)*

The extracted fit parameters are $A = (0.42 \pm 0.04)$ µV, the maximum $\Delta V_{\text{ANE}}$ amplitude, $\phi_0 = (4 \pm 5)°$, the phase shift, which expresses the angle between the MTJ's magnetization and $\nabla T$ when the temperature gradient is aligned parallel to $\mu_0 H$, and $V_0 = (0.00 \pm 0.026)$ µV, the offset voltage.

The small value obtained for $\phi_0$ indicates an excellent magnetization easy axis alignment with the external field direction. This also reveals, that when $\nabla T$ and $M$ are aligned parallel or antiparallel, $\Delta V_{\text{ANE}} = 0$. This corresponds to the angles $\phi_{\nabla T} = 0°$ and $\phi_{\nabla T} = 180°$, as indicated in the projection in figure 2 b). Both maximum amplitudes are located at $\phi_{\nabla T} = 90°$ and $\phi_{\nabla T} = 270°$, when $M$ and $\nabla T$ are perpendicular to each other. In conclusion, our findings are consistent with the cross product definition of the ANE. Besides this, the vanishing offset $V_0$ confirms the ANEs symmetry with respect to the magnetization direction.

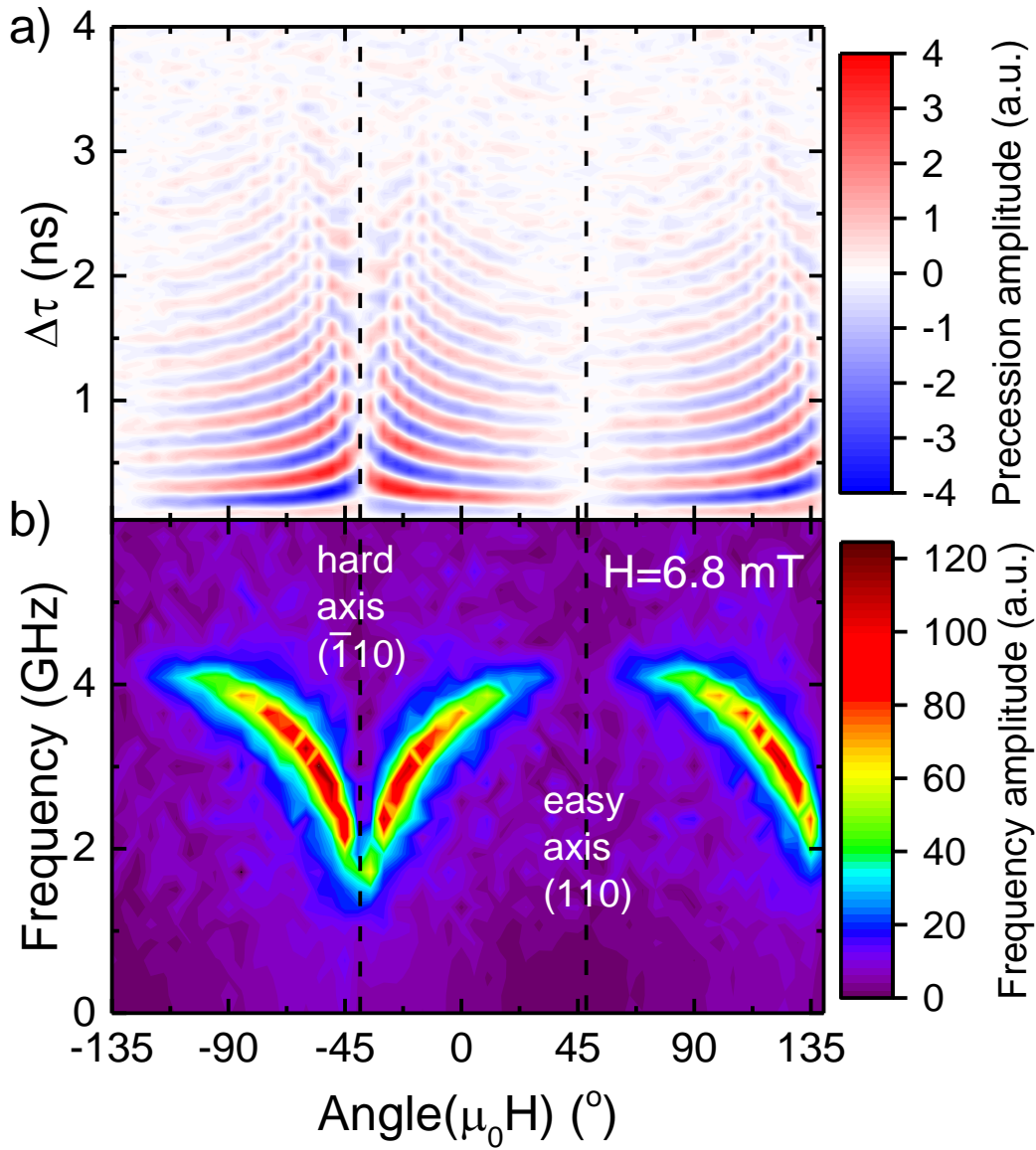

*Figure 3* **a)** Precessional dynamics from all-optical pump-probe experiments recorded for a CoFeB thin film, by rotating the $\mu_0 H = 6.8$ mT in the film plane in steps of 5°. The precession amplitude is coded from negative deflection (blue) to positive deflection (red). **b)** The precession frequencies extracted by FFT (different color code for a better distinction). The magnetic hard and easy axes are marked by the dashed black lines. The precession frequency increases towards the magnetic easy axis and declines towards the hard axis.

During the course of our measurements, we find that the MTJs possess an in-plane magnetic anisotropy. This is indicated in TMR as well as in MOKE measurements on CoFeB/MgO films. Additional magnetic anisotropy contributions could also influence the characteristics of the ANE effect.

In order to suppress these contributions to the voltage signal generated by in-plane temperature gradients, we first analyze the magnetic anisotropy for CoFeB thin films deposited on MgO substrates in detail. For this purpose, magnetization dynamics experiments, rotating $\mu_0 H$ in the sample plane are performed with an angle resolution of 5°, as depicted in figure 3 a). Here, the precessional dynamics on the nanosecond time scale are plotted vs. the rotation of $\mu_0 H$ in a false color plot, showing the negative/positive precession amplitude in blue/red. Figure 3 b) shows the precession frequencies extracted by fast Fourier transform. The frequency amplitudes are false color coded using a different color scheme for a better distinction. In accordance to the analysis presented in reference [33], we interpret our data as follows. The plot shows a declining precession frequency near the hard axis pointing in the $(\bar{1}10)$ direction and frequency increase when $\mu_0 H$ is rotated towards the easy axis pointing in the $(110)$ direction. The frequency reaches saturation and the amplitude declines in the vicinity of the easy axis, because the applied field amplitude ($\mu_0 H = 6.8\,\text{mT}$) is sufficient to saturate the sample, but too small to force the magnetization slightly out of the easy axis. This means, CoFeB grown on MgO exhibits a uniaxial magnetic anisotropy (UMA) with the easy axis along the $(110)$ crystalline direction.

Although the MTJ's elliptic shape is aligned with the vertex along the $(100)$ direction, for a 1.5 vertex/co-vertex ratio and a layer thickness in the nanometer range, the calculated demagnetizing fields due to shape anisotropy are approximately 2 mT, using geometrical considerations given in reference [34]. Therefore, solely the magneto-crystalline anisotropy remains as a significant factor leaving the easy axis in the $(110)$ direction. Taking those findings into account, the MTJ is placed with the easy magnetization axis parallel to the applied field $\mu_0 H$ for the ANE measurements.

In addition to this we also exclude contact resistance or bond wire geometry as an origin for this behavior, because repetition of those measurements with the contact wires attached at different angles to the magnetic field as well as at various positions and distances from the MTJ all return the same qualitative and quantitative characteristics (not shown here).

Finally, from these findings, the ANE coefficient can be estimated, considering that the CoFeB saturation magnetization is $M_S \approx 1.6\,\text{T}$ and the in-plane temperature difference $\Delta T \approx 8\,\text{K}$. The maximum $\Delta V_{\text{ANE}}$ value needs to be divided by two, because the shift in figure 2 a) influences the voltage measured in both parallel magnetization alignment directions. Starting with a homogeneous heating scenario, where the in-plane $\nabla T \approx 0$, also results in $\Delta V_{\text{ANE}} = 0$. However, an in-plane $\nabla T \neq 0$ shifts $V_{\text{p1}}$ to higher values, while it shifts $V_{\text{p2}}$ to lower values. Thus, the contribution to the ANE is given by $\frac{1}{2}\Delta V_{\text{ANE}}$. This results in an anomalous Nernst coefficient of $K_N = \frac{\frac{1}{2}\Delta V_{\text{ANE}}}{M_S \cdot \Delta T} \approx 1.6 \cdot 10^{-8}\,\frac{V}{TK}$. How does this value compare to previously published results? In 2014, Lee et al. determined the anomalous Nernst coefficients in ferromagnet/ non-magnet heterostructures for non-magnet materials with different spin hall angles in the range from $10^{-6}\,\frac{V}{TK}$ to $10^{-8}\,\frac{V}{TK}$ [23]. The value found in our detection scheme through magneto-Seebeck measurements, agrees well with the order of magnitude with these values. However, in their measurement the contributions from the ANE and the spin Seebeck effect are difficult to disentangle. A further look into literature reveals that $K_N$ varies between different materials by several orders of magnitude. For instance,

Wells et al. extracted an anomalous Nernst coefficient $K_\mathrm{N} = 2.3 \cdot 10^{-6} \frac{\mathrm{V}}{\mathrm{TK}}$ from measurements on perpendicularly (out-of-plane) magnetized amorphous CoFeB nanowires [31]. For FePt, Mizuguchi et al., and later Sakuraba et al. determined an anomalous Nernst coefficient of $\sim 0.5 \cdot 10^{-7} \frac{\mathrm{V}}{\mathrm{TK}}$ [35, 36]. A similar value of $\sim 1.3 \cdot 10^{-7} \frac{\mathrm{V}}{\mathrm{TK}}$ [19] was found by Weiler et al. for Ni. The comparison shows that our experimental method is extremely sensitive. We estimate that even for an anomalous Nernst coefficient as small as $10^{-9} \frac{\mathrm{V}}{\mathrm{TK}}$ a detection would be possible.

## 4. Conclusion

We investigated how in-plane temperature gradients in single magnetic tunnel junctions enhances or decreases the out-of-plane thermovoltage in TMS measurements. The extracted voltage shows a symmetric characteristic that can be clearly attributed to the anomalous Nernst effect with respect to the uniaxial magnetic anisotropy of the sample. This uniaxial magnetic anisotropy is verified by magnetization dynamics measurements.

Primarily we observe, that the ANE affects only the Seebeck voltage in the parallel magnetization alignment and the ANE voltages are two orders of magnitude smaller compared to the TMS voltages. Therefore, the influence on the overall TMS ratio needs to be considered in the analysis if in-plane temperature gradients are present, even if it is small. Nevertheless, the ANE can be clearly identified and extracted from TMS measurements of pseudo spin valve MTJs.

In the case of MTJs with one antiferromagnetically pinned and one switching electrode, the occurrence of ANE due to inhomogeneous heating and the presence of in-plane temperature gradients will lead to a deviation in the magneto-Seebeck voltage from the real value. However, in this configuration it is not possible to disentangle both contributions.

In our experiments, samples with different MgO barrier thicknesses are measured and show qualitatively similar characteristics. Thus, we can conclude that there is no significant influence of the MgO layer thickness on the ANE contribution.

Within this study, we illustrate the first detection of the anomalous Nernst effect in MTJs on such short length scales also obtaining a high spatial resolution. These results demonstrate very clearly the importance of homogenous laser heating to avoid unintended effects in case of TMS measurements by laser heating. The measurements show a clear dependence of the extracted ANE effect on the angle between the magnetization and the temperature gradient. Together with a proper calibration, and a combination of the investigated effects and technologies enables the construction of a direction dependent thermometer. This thermometer would not only sense the temperature, but also the direction of change. Besides this, the exponential temperature decay in the sample plane together with the sensitivity of this method, leave room for further device miniaturization beyond the micrometer scale.

## 5. Acknowledgements

The authors gratefully acknowledge financial support by the Deutsche Forschungsgemeinschaft (DFG) within the priority program SpinCaT (SPP 1538).

## 6. Methods

### Sample fabrication

The sample stack of the investigated thin films consists of Au 70 nm/ Ru 3 nm/ Ta 5 nm/ CoFeB 5.4 nm/ MgO 1.68 nm/ CoFeB 2.5 nm/ Ta 10 nm/ MgO (100) substrate. The CoFeB electrodes are fabricated by magnetron sputtering using 2 inch targets with a composition of $Co_{0.2}Fe_{0.6}B_{0.2}$ (analysis Co:Fe 0.32:0.68). In a separate chamber, the MgO barrier is e-beam evaporated without breaking the vacuum. The Ru capping layer is deposited by e-beam evaporation and prevents the underlying layers from oxidation. Ex-situ annealing with applied bias field is performed to crystallize the amorphous CoFeB electrodes and the MgO layer to obtain coherent interfaces and to activate the diffusion of B into the Ta layers [37–39]. Afterwards elliptical MTJs are patterned to a size of 6 μm × 4 μm with the long axis parallel to the direction of the magnetic field applied during the annealing by lithography processes. For thermal and electrical isolation, $Ta_2O_5$ is sputtered in the surroundings of the single MTJs. The Au layer pads on top are necessary to enable electrical contacting. A detailed description of the sample fabrication can be found in reference [11].

### Setup

#### Magneto-Seebeck experiment

For the generation of a temperature gradient across the layer stack, a laser diode (TOPTICA ibeam smart) with a wavelength of 638 nm and a maximum power of 150 mW is used. The laser is focused to a minimum diameter of ~2 μm full-width at half-maximum by utilizing a microscope objective (NIKON 20x, WD 20.5 mm). The generated thermovoltage is detected with a lock-in amplifier. The laser diode is modulated with a square wave at a frequency of 77 Hz, which is used as modulation frequency for the lock-in amplifier. For magnetization-dependent measurements, the sample is placed in between two pole shoes of an electromagnet. The implemented linear stages with motorized actuators for the horizontal (x-direction) and vertical (y-direction) movement enable an exact positioning of the laser beam on the sample surface together with a high spatial resolution of 0.2 μm. This setup allows the recording of the generated thermovoltage in z-direction depending on the magnetization direction by heating the sample at different positions over a defined area. In this study, the measured area is adapted to the junction size and in respect to the backlash of the actuators a dimension of 30 μm × 30 μm with a resolution of 1 μm is preferred.

#### Magnetization dynamics

The all-optical pump-probe Faraday configuration uses a 400 nm pump and 800 nm probe beam from a 1 kHz Ti:Sapphire laser system with 120 fs pulse lengths. The pump fluence is $F_{\text{pump}} = 5.7 \frac{\text{mJ}}{\text{cm}^2}$. The delay can be varied from $0 - 8$ ns. The sample is situated in a constant applied magnetic field which can be rotated in the sample plane.

### Temperature distribution simulations

The temperature distributions were obtained by finite element modelling with the software package COMSOL version 4.2a, including the heat transfer module. Most values for the necessary material parameters (specific heat $c$, thermal conductivity $\kappa$, density $\rho$) were taken from reference [7]. For $Ta_2O_5$ we assumed $c = 135.6 \frac{\text{J}}{\text{mol} \cdot \text{K}}$, $\kappa = 0.3 \frac{\text{W}}{\text{m} \cdot \text{K}}$, and $\rho = 8270 \frac{\text{kg}}{\text{m}^3}$ according to references [40, 41]. In contrast to the work presented in reference [7], here we implemented a fully three-dimensional model of the junction. The laser-heating was taken into account as a volumetric heating source $H \sim \exp\left(-\frac{z}{\lambda_{opt}} - \frac{2(x-x_0)^2+(y-y_0)^2}{w^2}\right)$, where $z = 0$ refers to the surface of the top Au electrode.

## Authors contributions

U.M., J.W. and M.M. designed and set up the experiments U.M. performed the TMS/ANE measurements. U.M. and T.H. prepared the samples. H.U. performed the COMSOL simulations, analyzed and discussed the temperature data with U.M. and J.W. R.R.T., C.L.C. and R.I.T. performed the magnetization dynamics experiments, analyzed and discussed the anisotropy data with U.M. and J.W. U.M. and J.W. analyzed the data and discussed the thermal effects with T.H., O.R. and T.K. U.M. and J.W. prepared the manuscript. All authors discussed the experiments and the manuscript. M.M. and A.T. coordinated the research.